\documentclass[fleqn,usenatbib]{mnras}

\usepackage{newtxtext,newtxmath}
\usepackage{lineno}
\usepackage[T1]{fontenc}

\DeclareRobustCommand{\VAN}[3]{#2}
\let\VANthebibliography\thebibliography
\def\thebibliography{\DeclareRobustCommand{\VAN}[3]{##3}\VANthebibliography}
\usepackage{bm}
\usepackage{graphicx}	
\usepackage{amsmath}	

\title[Scaling measurement of MHD turbulence]{Measurement of the scaling slope of compressible magnetohydrodynamic turbulence by synchrotron radiation statistics}

\author[Zhang et al.]{
Xue-Wen Zhang,$^{1}$
Jian-Fu Zhang,$^{1,2}$\thanks{E-mail: jfzhang@xtu.edu.cn}
Ru-Yue Wang $^{1}$
and Fu-Yuan Xiang$^{1,2}$\thanks{E-mail: fyxiang@xtu.edu.cn}
\\
$^{1}$Department of Physics, Xiangtan University, Xiangtan 411105, China,\\ 
$^{2}$Key Laboratory of Stars and Interstellar Medium, Xiangtan University, Xiangtan 411105, China
}

\date{Accepted XXX. Received YYY; in original form ZZZ}

\pubyear{2023}

\begin{document}
\label{firstpage}
\pagerange{\pageref{firstpage}--\pageref{lastpage}}
\maketitle

\begin{abstract}
Based on magnetohydrodynamic turbulence simulations, we generate synthetic synchrotron observations to explore the scaling slope of the underlying MHD turbulence. We propose the new $Q$-$U$ cross intensity $X$ and cross-correlation intensity $Y$ to measure the spectral properties of magnetic turbulence, together with statistics of the traditional synchrotron $I$ and polarization $PI$ intensities. By exploring the statistical behavior of these diagnostics, we find that the new statistics $X$ and $Y$ can extend the inertial range of turbulence to improve measurement reliability. When focusing on different Alfv{\'e}nic and sonic turbulence regimes, our results show that the diagnostics proposed in this paper not only reveal the spectral properties of the magnetic turbulence but also gain insight into the individual plasma modes of compressible MHD turbulence. The synergy of multiple statistical methods can extract more reliable turbulence information from the huge amount of observation data from the Low-Frequency Array for Radio astronomy and the Square Kilometer Array.
\end{abstract}


\section{Introduction}
The magnetized turbulent fluids in astrophysical environments can be usually described by magnetohydrodynamic (MHD) turbulence theory, which plays a critical role in many astrophysical processes such as star formation (\citealt{Mac2004}), heat conduction  (\citealt{Narayan2001}), magnetic reconnection (\citealt{Beresnyak2017}), and acceleration of cosmic rays  (\citealt{Yan2008,Zhang2021,Zhang2023}). Therefore, studying the properties of MHD turbulence helps to advance the theory of MHD turbulence and to understand astrophysical processes associated with MHD turbulence.

Here, we briefly describe three significant advances made in earlier research on turbulence. The first is about incompressible non-magnetized turbulence. By using a self-similarity assumption of the turbulence cascade, Kolmogorov (\citeyear{Kolmogorov1941}, henceforth K41) derived a power-law relation of $E(k) \sim k^{-5/3}$ in the inertial range, which is called a classic Kolmogorov spectrum. The second is about incompressible magnetized turbulence. Iroshnikov \& Kraichnan (\citeyear{Iroshnikov1963,Kraichnan1965},~henceforth IK65) obtained the power-law scaling of $E(k) \sim k^{-3/2}$ in the inertial range by introducing nonlinear energy cascade. Although IK65 advanced the K41 theory by considering the effect of magnetic fields, it ignored a critical issue that the turbulence should be anisotropic in the magnetized fluids (\citealt{Montgomery1981}). The third is still about incompressible magnetized turbulence but focuses on the anisotropy of MHD turbulence due to turbulent magnetic fields. Focused on the nonlinear energy cascade of incompressible  strong MHD turbulence, Goldreich \& Sridhar   (\citeyear{Goldreich1995}, hereafter GS95) provided the theoretical predictions on the power-law scaling and anisotropic relationship, as described in more detail in Section \ref{MHDtheory}.

At present, many numerical simulations have significantly increased our knowledge on the scaling, anisotropy and compressibility of MHD turbulence (e.g., \citealt{Cho2002}; see textbook by \citealt{Beresnyak2019}; and a recent review by \citealt{Beresnyak2019LRCA} ). 
The properties obtained by the simulation of MHD turbulence can understand the acceleration and propagation of cosmic rays (\citealt{Yan2002}).  
In particular, the turbulent reconnection model proposed in \citeauthor{Lazarian1999} (\citeyear{Lazarian1999}, hereafter LV99), which provides a new interpretation for GS95 theory from the perspective of eddies, has been applied to various astrophysical environments such as gamma-ray bursts (\citealt{Petrosian2003,Zhang2011}), microquasars (\citealt{de2005}), active galactic nuclei (\citealt{Kadowaki2015}) and radio galaxies (\citealt{Brunetti2016}).

Due to the large scale of astrophysical system with a high Reynolds number $R_{\rm e} > 10^{10}$, it is challenging to simulate a realistic astrophysical environment by direct numerical simulation. The currently available 3D MHD simulations can achieve the case of the Reynolds number $R_{\rm e}\simeq10^5$ (e.g., \citealt{Beresnyak2019}). A distinctive feature is that the realistic inertial range in astrophysical turbulence is much greater than that revealed by numerical simulations. It is more effective to move away from direct numerical simulations and develop statistical techniques using observational data, to explore the properties of MHD turbulence.

When relativistic electrons motion in turbulent magnetic fields, they produce synchrotron radiation fluctuations providing information on the magnetic fields (\citealt{Schnitzeler2007,Lazarian2012,Lazarian2016} henceforth LP12 and LP16, \citealt{Iacobelli2013,VanEck2017,West2020,Sun2022} ). Based on the modern understanding of MHD turbulence and synchrotron radiation theory, LP12 explored the properties of MHD turbulence by statistics of synchrotron total intensity fluctuations. They predicted that synchrotron intensity fluctuations are anisotropic with a long extension along the direction of magnetic fields. Using the ratio between quadrupole and monopole components can determine the anisotropy of MHD turbulence, which is sensitive to the compressibility of underlying turbulence. These theoretical predictions on anisotropy have been confirmed successfully using numerical simulations (\citealt{Herron2016,Lee2019,Wang2022}). These studies are opening new avenues for exploring MHD turbulence using observational data.

Moreover, LP16 proposed to recover the properties of MHD turbulence using synchrotron polarization intensity fluctuations (including Faraday rotation fluctuations). They developed two main techniques from the perspective of analytical theory, i.e., polarization frequency analysis and polarization spatial analysis. The former, using a variance of synchrotron polarization intensity (or its derivative) as a function of the square of the wavelength, was tested by \cite{Zhang2016}. The latter, making use of spatial correlations of synchrotron polarization intensity (or its derivative) at the fixed wavelength as a function of the spatial separation $R$, was tested by \cite{Lee2016} and \cite{Zhang2018}. As confirmed, these two methods obtain the scaling index of the underlying turbulence cascade in the inertial range. Compared with synchrotron radiation, polarized radiation can reveal not only information about magnetic fields in the plane of the sky but also that parallel to the line of sight (LOS).

From the perspective of synthetic observations, numerical dissipation inevitably limits the extension of the power-law range, and the greater the inertial range, the higher the reliability of the measurement. However, from an observational point of view, the scaling index measurement of MHD turbulence is also limited by the telescope's resolution and data noise. It is necessary to synergize multiple techniques to reveal the properties of MHD turbulence and enhance the reliability of the measurement results.

This paper aims to advance the study of the power-law scaling properties of MHD turbulence. With the power spectrum (PS) and structure function (SF) methods for synchrotron diagnostics, we propose two new statistical quantities to explore the scaling slope properties of compressible MHD turbulence. The paper is structured as follows. Section~\ref{theo_des} describes theory aspects involving the basic theory of MHD turbulence, synchrotron radiative process and statistical methods. Section~\ref{data_produce} introduces the setups of the numerical simulation of MHD turbulence. Sections~\ref{results1} and \ref{results2} present the numerical results. In Sections~\ref{disc} and \ref{sum}, we provide our discussion and summary, respectively.

\section{Theoretical description}
\label{theo_des}

\subsection{MHD turbulence theory}
\label{MHDtheory}

GS95 theory is generally considered the basis for MHD turbulence. Note that GS95 theory focused on incompressible strong MHD turbulence with Alfv{\'e}nic Mach number $M{\rm_A}=V_{\rm L}/V_{\rm A}\simeq 1$, where $V_{\rm L}$ is the injection velocity at the injection scale $L_{\rm inj}$ and $V_{\rm A}$ is the Alfv{\'e}nic velocity. 
This theory combined the motions of the eddies perpendicular to the magnetic field with those parallel to the magnetic field by the critical balance condition $l_{\perp}/v_{\perp}=l_{\parallel}/V_{\rm A}$, where 
$v_{\perp}$ is the velocity at the scale $l$, and the scales $l_\parallel$, $l_\perp$ represent the parallel and perpendicular scales of eddies, respectively.
They found that the motions of eddies perpendicular to the magnetic field have similar properties to Kolmogorov turbulence with the spectrum of $E(k_{\perp})\propto {\epsilon}^{2/3}k_{\perp}^{-5/3}$, and the velocity-scale relation of $v_{\perp}\propto(\epsilon l_{\perp})^{1/3}$, where $k_{\perp}$ is the wave-vector component perpendicular to the magnetic field and $\epsilon$ is the rate of energy cascade. According to the velocity-scale relation and critical balance condition, they predicted an anisotropic relationship of
\begin{equation}
 \label{scale_perp_parallel}
 l_{\parallel}\sim V_{\rm A}{\epsilon ^{-1/3}}l_{\perp}^{2/3},
\end{equation}
which delineates the dependencies between the perpendicular and parallel scales of the eddies. 

Later, the GS95 theory was generalized from the trans-Alfv{\'e}nic turbulence to sub-Alfv{\'e}nic and super-Alfv{\'e}nic ones, respectively (LV99; \citealt{Lazarian2006}). For the former, $M_{\rm A}<1$, that is, the turbulence drives with the injection velocity $V_{\rm L}$ less than the Alfv{\'e}nic velocity $V_{\rm A}$, LV99 found that the turbulence cascade corresponds to two regimes. The first regime is a weak turbulence cascade ranging from the injection scale $L_{\rm inj}$ to the transition scale $L_{\rm tr}=L_{\rm inj}M_{\rm A}^2$. The second one is strong turbulence from the transition scale $L_{\rm tr}$ to the dissipation scale $L_{\rm diss}$, where the energy cascade perpendicular to the magnetic field is analogous to the hydrodynamic Kolmogorov cascade. In this strong turbulence regime, they derived the turbulence velocity as
\begin{equation}
 \label{sub_alfven_velocity}
 v_{\perp} \approx V_{\rm L}L_{\rm{ inj}}^{-1/3}M_{\rm A}^{1/3}l_{\perp}^{1/3},
\end{equation} 
and the anisotropic relation as 
\begin{equation}
 \label{sub_alfven_scale}
 l_{\rm \parallel}\approx L_{\rm inj}^{1/3}M_{\rm A}^{-4/3}l_{\perp}^{2/3}.
\end{equation} 
When taking $M_{\rm A}=1$, the above equations will return to the relevant expressions of GS95 theory.

As for the latter, $M_{\rm A}>1$, the MHD turbulence starting from the injection scale $L_{\rm inj}$ is almost no constraint of the magnetic field and has properties similar to those of hydrodynamic turbulence. With the cascade of turbulence, it experiences a transition from hydrodynamic-like turbulence to MHD one at the scale $L_{\rm A}=L_{\rm inj}M_{\rm A}^{-3}$. However, from the scale $L_{\rm A}$ to $L_{\rm diss}$, the turbulence again follows the characteristics of GS95 theory, having the velocity-scale relation of
\begin{equation}
 \label{sup_alfven_velocity}
 v_{\rm \perp} \approx V_{\rm L}L_{\rm inj}^{-1/3}l_{\perp}^{1/3},
\end{equation}
and the anisotropy of 
\begin{equation}
 \label{anis_l}
 l_{\parallel}\approx L_{\rm inj}^{1/3}M_{\rm A}^{-1}l_{\perp}^{2/3}.
\end{equation}

At present, the properties of compressible MHD turbulence have become an important part of the modern understanding of MHD turbulence theory. Compressible MHD turbulence can be decomposed into three modes, namely Alfv{\'e}n, slow and fast modes, as confirmed by numerical simulations (\citealt{Cho2002,Cho2003,Kowal2010}). 
Specifically, they found that Alfv{\'e}n and slow modes follow the GS95-type scaling law, namely $E(k_{\perp})\propto k_{\perp}^{-5/3}$ and the scale-dependent anisotropy, while fast mode presents the scaling law of $E(k_{\perp})\propto k_{\perp}^{-3/2}$ and the isotropy. In addition, for compressible MHD turbulence, the Alfv{\'e}n mode is incompressible, while the slow and fast modes, called magnetosonic modes, are compressible. \footnote{ When focusing on the compressible MHD turbulence as done in this work, one, for the sake of simplicity, can call the slow mode as a compressible mode. However, for the incompressible MHD turbulence with the plasma parameter $\beta\gg 1$, the slow mode is a pseudo-Alfv\'en mode with a purely solenoidal feature.}

Despite the progress made in the development of MHD turbulence theory, there are still a lot of controversial issues. For example, \cite{Maron2001} numerically studied the incompressible MHD turbulence and found a shallow energy spectral index of $k^{-3/2}$ different from $k^{-5/3}$ given by GS95. Subsequently, to explain this shallow index, \cite{Boldyrev2006} proposed the dynamic alignment model to modify the GS95 scaling index from $-5/3$ to $-3/2$. Later, \cite{Beresnyak2010} and \cite{Beresnyak2014} thought that the spectral index $-5/3$ cannot extend to the entire inertial range, but deviate near the part of the injection scale (see also \citealt{Beresnyak2019} for the recent review). This can explain why the low-resolution numerical simulations generate a shallower spectral index, while the results of higher-resolution numerical simulations are consistent with GS95. However, some recent studies, e.g., \cite{Chandran2015}, agreed with the dynamical alignment theory.

By analyzing the power spectrum of super-sonic turbulence from the solenoidal and compressive driving ways, \cite{Federrath2013MNRAS} found the velocity spectral indices satisfy with $k^{-2}$. In the case of solenoidal driving, the spectrum of the density-weighted velocity $\rho^{1/3}v$ satisfies with $k^{ -1.74}$, while in the case of compressive driving, the slope is significantly steeper and close to $k^{-2.1}$. This result is consistent with the compressible turbulence theory (\citealt{Galtier2011}), which predicts the scaling of density-weighted velocity $k^{ -19/9}$. Recently, \cite{Mallet2017} proposed the intermittency model to modify MHD turbulence theory at scales close to the dissipation scales. However, because of the limitations of numerical simulations, it is difficult to confirm. 

Until now, many attempts have not significantly changed the framework of the GS95 theory. Although our study below is based on the GS95 theory, the change, in theory, does not affect our results based on synthetic synchrotron observations.

\subsection{Synchrotron emission fluctuations} 
For the sake of simplicity, this work assumes that relativistic electrons interacting with the turbulent magnetic field satisfy a homogeneous and power-law energy distribution of $N(E)= N_0E^{-p}$, where $p$ and $E$ represent the spectral index and energy of relativistic electrons, respectively. Here, $N_0$ is the normalization constant of electrons. 
According to the classic textbooks (\citealt{Rybicki1979,Longair2011}), observable Stokes parameters under the condition of no Faraday rotation effect can be expressed as follows (see also, e.g.,\citealt{Waelkens2009}; LP16):
\begin{equation}
 \label{stokes_I0}
{I}({\bm X})=\int_{0}^{L} dz(B_{\rm x}^2({\bm x})+B_{\rm y}^2({\bm x}))^{\frac{p-3}{4}} (B_{\rm x}^2({\bm x})+B_{\rm y}^2({\bm x})),
\end{equation}
\begin{equation}
\label{stokes_Q0}
Q_{0}({\bm X})=\int_{0}^{L} dz(B_{\rm x}^2({\bm x})+B_{\rm y}^2({\bm x}))^{\frac{p-3}{4}} (B_{\rm x}^2({\bm x})-B_{\rm y}^2({\bm x})),
\end{equation}
\begin{equation}
\label{stokes_U0}
U_{0}({\bm X})=\int_{0}^{L} dz(B_{\rm x}^2({\bm x})+B_{\rm y}^2({\bm x}))^{\frac{p-3}{4}} (2B_{\rm x}({\bm x}) B_{\rm y}({\bm x})),
\end{equation}
where $L$ is the integral depth along the LOS, $B_{\rm x}$ and $B_{\rm y}$ the components of the magnetic field perpendicular to the LOS, and ${\bm X}=({x},{y})$ the spatial coordinate in the plane of the sky.

Focusing on linear polarization synchrotron radiation, we have a complex vector 
\begin{equation}
\label{obsrved_polarization}
 {\bm P}({\bm X}, \lambda^2)=Q+iU=\int_0^{L}d{z}P_{\rm in}({\bm X}, z)e^{2i {\rm \phi}({\bm X}, z)},
\end{equation}
describing polarization states in the plane of the sky. In this equation, the exponential factor involves Faraday rotation effect. The observed polarization angle $\phi$ is expressed by
\begin{equation}
 \label{polarization_angle}
\phi = \phi_{0} + {\lambda^2} \rm RM,
\end{equation}
where the angle $\phi_{0}$ is the intrinsic angle. The Faraday rotation measure $\rm RM$ is written as
 \begin{equation}
 \label{rotation_measure}
{\rm RM}(\bm{X}, z)=0.81 \int_{0}^{z}n_{\rm e}({\bm X}, z')B_{\parallel}({\bm X}, z')dz'~ {\rm rad} ~{\rm m^{-2}},
\end{equation}
where $n_{\rm e}$ is the density of thermal electrons, $B_{\parallel}$ the component of the magnetic field along the LOS. The integral length $L$ is along the LOS from the position of the source at $z$ to the observer. Moreover, the part $P_{\rm in}$ of integrated function in
Equation (\ref{obsrved_polarization}) represents the intrinsic polarization intensity density and can be expressed by $P_{\rm in}\equiv(Q_{0}, U_{0})$. After including Faraday rotation effect, the new Stokes parameters $Q$ and $U$ can be rewritten as 
\begin{equation}
 \label{stokes_Q}
Q({\bm X},{\rm \lambda^2})=Q_{\rm 0} {\rm \cos2\phi} + U_{\rm 0} {\rm \sin2\phi},
\end{equation}
\begin{equation}
 \label{stokes_U}
U({\bm X},{\rm \lambda^2})=U_{\rm 0}{\rm \cos2\phi} - Q_{\rm 0}{\rm \sin2\phi},
\end{equation}
from which we obtain the synchrotron polarization intensity of
\begin{equation}
\label{polarization_intensity}
PI=\sqrt{Q^2+U^2}.
\end{equation}

A complete description of synchrotron radiation can be encoded by a polarization matrix, e.g., as done in Equation (E1) of LP12. This paper focuses on the correlation statistics between $Q$ and $U$. The first is the $Q$-$U$ cross intensity defined by \footnote{ This definition was used to explore the anisotropy of MHD turbulence by the structure function (LP12) and to trace magnetic field directions by gradient techniques (\citealt{Lazarian2018}).}
\begin{equation}
X^2=QU, \label{X2}
\end{equation}
 which is related to the relative importance of $Q$ and $U$. In general, both the different turbulence properties and the level of Faraday rotation depolarization will lead to different $Q$ and $U$ values, and different ratios of $Q$ and $U$.  
As done in LP16, when considering the correlation function of the polarization complex vector ${\bm P}$, we have 
\begin{multline}
\label{polarization_corr}
\langle P(\bm{X_1})P^*(\bm{X_2})\rangle=\langle Q(\bm{X_1})Q(\bm{X_2})+ U(\bm{X_1})U(\bm{X_2})\rangle\\ +i\langle U(\bm{X_1})Q(\bm{X_2})- Q(\bm{X_1})U(\bm{X_2})\rangle,
\end{multline}
which is split into real and imaginary parts that are separately invariant with respect to frame rotation. The symmetric real part carries the most straightforward information about the magnetized turbulent medium and has been numerically studied in \cite{Zhang2016}. LP16 predicted that the antisymmetric imaginary part reflects helical correlations of the magnetic field, which still needs numerical testing. In addition, analytical studies demonstrated that the anisotropy of the MHD turbulence can generate the observable antisymmetric correlations. Based on the antisymmetric imaginary part in Equation (\ref{polarization_corr}), we rewrite the cross-correlation intensity as
\begin{equation}
 \label{technique_Y}
Y^2(\bm X')=\int{d^2{\bm X} [U({\bm X})Q({\bm X}+\bm X')-Q({\bm X})U({\bm X}+\bm X')]
}.
\end{equation}
Adopting Equations (\ref{X2}) and (\ref{technique_Y}),
we will explore the scaling property of MHD turbulence by comparing the traditional $PI$ and $I$ statistics. Note that the cross-correlation intensity $Y$ is covariant variable during rotation and translation transformation in the Stokes frame, while the cross-intensity $X$ are unchanged only when the Stokes frame is translated.

\subsection{Statistical methods}\label{StatMeth}
Although turbulence is a complex and chaotic process, it allows us to use statistical methods to reveal its underlying properties. In this paper, we focus on the SF and PS methods. We first consider the simplest and often used correlation function for an arbitrary 2D physical quantity $\zeta$. According to the textbook by \cite{Monin1975}, correlation and structure functions are written as 
\begin{equation}
 \label{two_point_cf}
 {\rm CF}({\bm R})=\langle \zeta{({\bm X}+{\bm R})}\zeta({\bm X})\rangle,
\end{equation}
\begin{equation}
 \label{two_point_sf}
 {\rm SF}({\bm R})=\langle(\zeta{({\bm X}+{\bm R})}-\zeta({\bm X}))^2\rangle,
\end{equation}
and they satisfy the following relation
\begin{equation}
 \label{two_point_cfsf}
 {\rm SF}({\bm R}) = 2[{\rm CF(0)-CF}(\bm R)],
\end{equation}
where ${\bm R}$ is a separation vector, and $\langle...\rangle$ represents the average through the whole volume space. 

The PS, a common statistical tool in the study of turbulence, can provide information on the energy cascade of MHD turbulence, such as the spectral shape and index, the source and the sink. The PS of a two-dimensional physical quantity is expressed by 
\begin{equation}
\label{power_spectrum}
 P_{\rm 2D}({\bm K})=\frac{1}{(2\pi)^2}\mathrm{\int} \langle \zeta({\bm X})\zeta({\bm X}+{\bm R})\rangle  e^{{-i {\bm K}} \cdot  {\bm R}}d{\bm R}
\end{equation}
by the Fourier transform of the correlation function. The ring-integrated 1D spectrum for a 2D variable follows 
\begin{equation}
\label{Epower_spectrum}
 E_{\rm {2D}}({K})=\int_{K - 0.5}^{K + 0.5} P_{\rm {2D}}(K)d{K}.
\end{equation}
Note that there is a direct connection between PS and SF by the scaling slope: $E_{\rm 2D}(K) \propto K^{-m}$ and ${\rm SF}(R) \propto R^{m-1}$ (LP12; see also numerical confirmation in \citealt{Lee2016, Zhang2018}), where $m$ is equal to 8/3 for Kolmogorov power spectrum in two dimensions.

\begin{table*}
 \caption{The information of data cubes arising from the simulation of compressible MHD turbulence. Relevant parameters used to characterize data cubes are given as ---- $B_0$: mean magnetic field along the $x$ coordinate; $\beta$: plasma parameter; $L_{\rm tr}$: transition scale of strong turbulence in sub-Alfv\'enic regime; $L_{\rm A}$: transition scale of strong turbulence in the super-Alfv\'enic regime.}
 \label{tab:1}
  \begin{tabular}{lcccccccc}
  \hline
  Run &$B_{0}$&$M_{\rm A}$&$M_{\rm s}$&$\beta=2M^2_{\rm A}/M_{\rm s}^2$&$L_{\rm inj}$[2<$k$<3]&$L_{\rm inj}[k=2.5]$&$L_{\rm tr}$($L_{\rm A}$)[2<$k$<3] &$L_{\rm tr}$($L_{\rm A}$)[$k$ = 2.5]  \\
  \hline
 1 & 1.00 & 0.70 & 0.87 & 1.30 & [170.6, 256.0] & 204.8 & [83.59, 125.44] & 100.35\\[2pt]
 2 & 1.00 & 0.55 & 4.46 & 0.03 & [170.6, 256.0] & 204.8 & [51.61,\; 77.44] & 61.95 \\[2pt]
 3 & 1.00 & 0.65 & 0.48 & 3.67 & [170.6, 256.0] & 204.8 & [72.09, 108.16] & 86.53 \\[2pt]
 4 & 0.10 & 1.69 & 3.11 & 0.60 & [170.6, 256.0] & 204.8 & [35.34, \; 53.04] & 42.43 \\[2pt]
 5 & 0.10 & 1.72 & 0.45 & 29.30 & [170.6, 256.0] & 204.8 & [33.53, \; 50.31] & 40.25 \\[2pt] 
  \hline
 \end{tabular} 
\end{table*}

\section{MHD turbulence simulations}
\label{data_produce}
To generate synchrotron observations, we use a third-order accurate hybrid, essentially non-oscillatory (ENO) scheme (\citealt{Cho2002}) to solve ideal isothermal MHD equations in a periodic box of size $2\pi$:
\begin{equation}
\label{continuity}
\frac{\partial\rho}{\partial t} + \nabla\cdot (\rho {\bm v}) = 0, 
\end{equation}
\begin{equation}
\label{momentum_conservation}
\rho[\frac{\partial {\bm v}}{\partial t}+({\bm v}\cdot\nabla){\bm v}]+\nabla p-{\bm J} \times \frac{\bm B}{4\pi}= {\bm f},
\end{equation}
\begin{equation}
\label{magnetic_induction}
\frac{\partial{\bm B}}{\partial t} -  \nabla\times({\bm v}\times{\bm B}) = 0,
\end{equation}
\begin{equation}
\label{divergence}
\nabla \cdot {\bm B}
 = 0, 
\end{equation}
where $\rho$ is the density, 
$p=c^2_{s}\rho$ the thermal gas pressure, $\bm{J}$ the current density, and $\bm{f}$ an external driving force. The turbulence is driven by a solenoidal driving force at the wave number of $2<k<3$ (corresponding to the mean wavenumber of $k\approx 2.5$) (\citealt{Cho2003}).
By setting the initial mean magnetic field (along the $x$ axis) and the gas pressure, we run several simulations with the resolution of $512^3$ covering different MHD turbulence regimes. When the running reaches a statistically steady state, from output data cubes of the magnetic field, velocity, and density, we calculate the Alfv{\'e}nic and sonic Mach numbers to characterize the properties of each simulation. Specifically, Alfv{\'e}nic Mach number is obtained by $M_{\rm A}= \langle\bm V_{\rm L} /\bm V_{\rm A}\rangle$ and sonic Mach number by $M_{\rm s}=  \langle\bm V_{\rm L} /c_{\rm s}\rangle $, where $c_{\rm s}=\sqrt{p/\rho}$ is sound speed, and $V_{\rm A}\approx \frac{\bm B}{\sqrt{\rho}}$ is Alfv{\'e}nic velocity. The resulting numerical values are listed in Table \ref{tab:1}.

The compressible MHD turbulence can be decomposed into Alfv{\'e}n, slow and fast modes by using the following theoretical procedures (\citealt{Cho2002}; \citealt{Cho2003})  
 \begin{equation}
 \label{slow_vector}
\hat{\xi}_{\rm s}  \propto (1 + \frac{\beta}{2} - \sqrt{D})(k_{\perp}{\hat{\bm k}}_{\perp}) +(-1 + \frac{\beta}{2} - \sqrt{D})(k_{\parallel}{\hat{\bm k}}_{\parallel}),  
\end{equation}
\begin{equation}
\label{fast_vector}
\hat{\xi}_{\rm f} \propto (1 + \frac{\beta}{2} +\sqrt{D})(k_{\perp}{\hat{\bm k}}_{\perp}) +(-1 + \frac{\beta}{2} + \sqrt{D})(k_{\parallel}{\hat{\bm k}}_{\parallel}),  
\end{equation}
\begin{equation}
\label{alfven_vector}
\hat{\xi}_{\rm A} \propto  -{\hat{\bm k}}_{\perp} \times {\hat{\bm k}}_{\parallel}, 
\end{equation}
in the Fourier space, where $D=(1+\frac{\beta}{2})^2-2\beta\cos^2 \theta$, and $\cos\theta = {\hat{\bm k}}_{\parallel}\cdot{\hat{\bm B}}_{0}$. By projecting the magnetic field into the displacement vectors $\hat{\xi}_{\rm f}$, $\hat{\xi}_{\rm A}$ and $\hat{\xi}_{\rm s}$, we get the individual components of three modes for magnetic field and then convert them to volume space by the Fourier inverse transform. Later, \cite{Kowal2010} proposed an optimized decomposition method by introducing a discrete wavelet transform, and decomposing each component of the magnetic fields into orthogonal wavelets using a discrete wavelet transform. This method, which depends on the local magnetic field rather than the mean magnetic field, is universal for decomposition in the super-Alfv{\'e}nic turbulence with weak magnetic fields. Since only sub-Alf\'enic turbulence is involved in our work when decomposing plasma modes, we use the Fourier decomposition method.

When exploring below the influence of the angle between the mean magnetic field and LOS on the power spectrum, we adopt the Euler rotation algorithm to rotate data cubes (\citealt{Lazarian2018,Carmo2020,Wang2022,Malik2023}). The components of the rotation matrix $\bm{\hat{F}}=\hat{\bm F}_x\hat{\bm F}_y\hat{\bm F}_z$ are expressed as follows:
\begin{equation}
{\bm{\hat{F}_x}}=
\left[\begin{array}{cccc}
1 &         0          & 0                   &\\
0 & \rm cos(\varphi_x) & -\rm sin(\varphi_x) & \\
0 & \rm sin(\varphi_x) & \rm cos(\varphi_x)  &
\end{array}
\right],
\end{equation}
\begin{equation}
{ \hat{\bm F}_y}=
\left[\begin{array}{cccc}
\rm cos(\varphi_y)  & 0 & \rm sin(\varphi_y) &\\
0 & 1 & 0 & \\
-\rm sin(\varphi_y) & 0 & \rm cos(\varphi_y) &\\
\end{array}
\right],
\end{equation}
\begin{equation}
{ \hat{\bm {F}}_z}=
\left[\begin{array}{cccc}
\rm cos(\varphi_z) & -\rm sin(\varphi_z) & 0 &\\
\rm sin(\varphi_z) & \rm cos(\varphi_z)  & 0 & \\
0 & 0 & 1 &\\
\end{array}
\right],
\end{equation}
where $\varphi_{m=x,y,z}$ is the rotation angle along the $x$, $y$, $z$ axis, respectively. For data cube $\Re(\bm r)$ of components of magnetic filed, the rotated data cube is obtained by ${\hat {\bm F} \Re({\bm \hat {\bm F}}^{-1}} {\bm r})$ transformation. Since the rotation of the cube is equivalent to the rotation of the observation frame in the opposite direction, we perform an inverse transformation of the position vector ${\bm{ r}}$.

\begin{figure*}
\centering
\includegraphics[width=1.8\columnwidth,height=0.5\textheight]{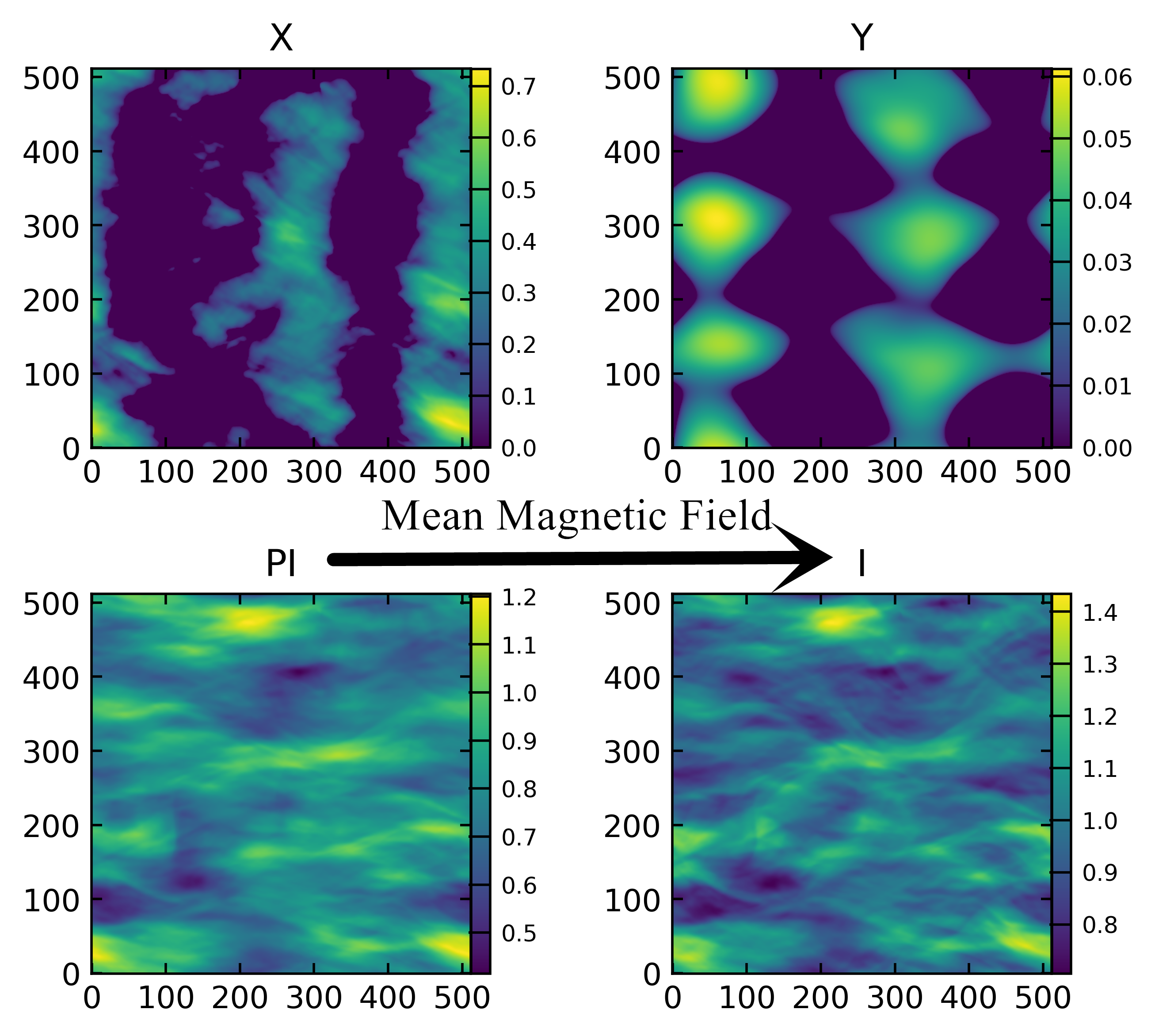}
\caption{Images of synchrotron radiation diagnostics: $Q$-$U$ cross ($X$), cross-correlation ($Y$), linear polarization $(PI)$ and total ($I$) intensities. The calculations without the Faraday rotation effect are based on run3 listed in Table \ref{tab:1}. The mean magnetic field is along the horizontal direction. }
\label{fig:Iamg_four}
\end{figure*}

\section{Statistical Results: Compressible MHD Turbulence}
\label{results1}
When generating synthetic synchrotron observations, we adopt dimensionless units. In the case of involving the Faraday rotation effect, we quantify relevant physical quantities: magnetic fields as $B=1.23 \rm\ \mu G$, thermal electron density as $n_{\rm e}=0.01 \rm\ cm^{-3}$, and the box size as $L=100 \rm\ pc$, which correspond to the Galactic ISM environment. In addition, we set the spectral index of relativistic electrons to be $p=2.5$.{\footnote{ Electron spectral indices are associated with a specific acceleration mechanism. For instance, \cite{de2005} predicted that turbulent reconnection acceleration provided a steeper spectral index $p=2.5$, confirmed numerically by \cite{Zhang2023}. The spectral index determined from observations will change in different astrophysical environments. Our previous studies demonstrated that the change in spectral indices could not impede the turbulence measurement using the synchrotron polarization technique (see LP12 for theoretical predictions; \citealt{Lee2016} and \citealt{Zhang2018} for numerical confirmations).}}

\subsection{Slope measurement of MHD turbulence without Faraday rotation}
Before performing statistical analysis, let us illustrate the map structures of the statistical diagnostics considered in this paper. Based on run3 listed in Table~\ref{tab:1}, we obtain the intensities of different synchrotron radiation diagnostics, namely $Q$-$U$ cross $X$, cross-correlation $Y$, linear polarization $PI$ and total $I$ intensities. Here, we do not consider the Faraday rotation effect. The imaging of these diagnostics is plotted in Figure~\ref{fig:Iamg_four} (using the real part of $X$, $Y$ to exhibit their maps), from which we can see that map structures of the $Q$-$U$ cross $X$, cross-correlation $Y$ are elongated along the direction perpendicular to the mean magnetic field while maps of linear polarization $PI$ and total $I$ intensities have nearly similar structures extending along the horizontal direction, i.e., the direction of the mean magnetic field. The perpendicular distribution from $X$ and $Y$ should be dominated by the Stokes parameter $U$, while the horizontal structure from $PI$ and $I$ is dominated by the Stokes parameter $Q$. In general, map structures in the Stokes parameter $Q$ are aligned with the direction of the mean magnetic field, while those of $U$ are perpendicular to the mean magnetic field. Moreover, the intensities of $PI$ and $I$ diagnostics have larger amplitudes than those of $X$ and $Y$. This is caused by amplitude changes in $Q$ and $U$, the intensity of which is associated with the magnetic field strength in the plane of the sky (see Equations (\ref{stokes_Q0}) and (\ref{stokes_U0})).

\begin{figure*}
\centering
\includegraphics[width=2\columnwidth,height=0.6\textheight]{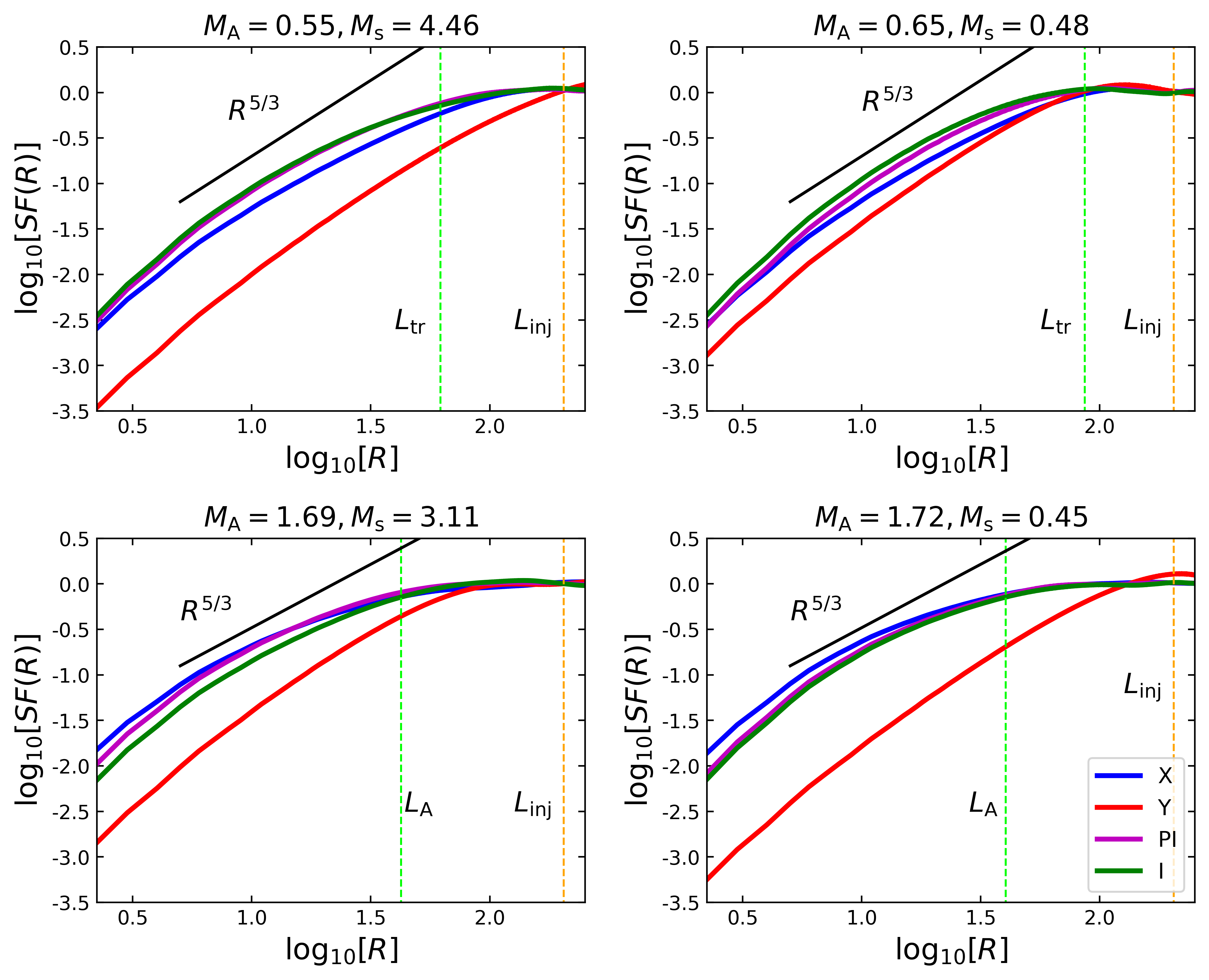}
 \caption{ Structure function of synchrotron radiation diagnostics: $Q$-$U$ cross ($X$), cross-correlation ($Y$), linear polarization $(PI)$ and total ($I$) intensities in different turbulence regimes. The yellow and green vertical dashed lines represent the injection and transition scales, respectively.}
 \label{fig:two_sf}
\end{figure*}

The SFs of $X$, $Y$, $PI$, and $I$ are plotted in Figure~\ref{fig:two_sf} for four different turbulence regimes: sub-Alfv{\'e}nic and supersonic (left upper panel), sub-Alfv{\'e}nic and subsonic (right upper), super-Alfv{\'e}nic and supersonic (left lower) and super-Alfv{\'e}nic and subsonic (right lower). As shown, SFs cannot recover the scaling of MHD turbulence in the regime ranging from the injection scale $L_{\rm inj}$ to the transition scale $L_{\rm tr}$ for $M_{\rm A} <1$ (or $L_{\rm A}$ for $M_{\rm A} >1$). These numerical results are in agreement with theoretical predictions of MHD turbulence cascade due to weak turbulent interaction (LV99; \citealt{Lazarian2006}). At the scale less than the transition scale $L_{\rm tr}$ (or $L_{\rm A}$), i.e., in the strong turbulence regime, these four diagnostics present the power-law distributions predicted by LV99 and \cite{Lazarian2006}. From the figures, We can see that: (1) in the case of sub-Alfv{\'e}nic turbulence (upper panels), the measurements from $X$ and $Y$ are closer to the slope index 5/3 than those from $PI$ and $I$; and (2) in the case of super-Alfv{\'e}nic turbulence scenario (lower panels), the $Y$ statistics can better determine the slope index 5/3 compared with the other three statistics $X$, $PI$ and $I$. Comparing sub-Alfv{\'e}nic and super-Alfv{\'e}nic turbulence, we find that super-Alfv{\'e}nic turbulence has a shorter inertial range for $X$, $PI$, and $I$. Interestingly, we find that statistics $Y$ can well reflect the scaling of 5/3 with a wide inertial range and does not depend on specific turbulence properties.

\begin{figure*}
\centering
 \includegraphics[width=2\columnwidth,height=0.6\textheight]{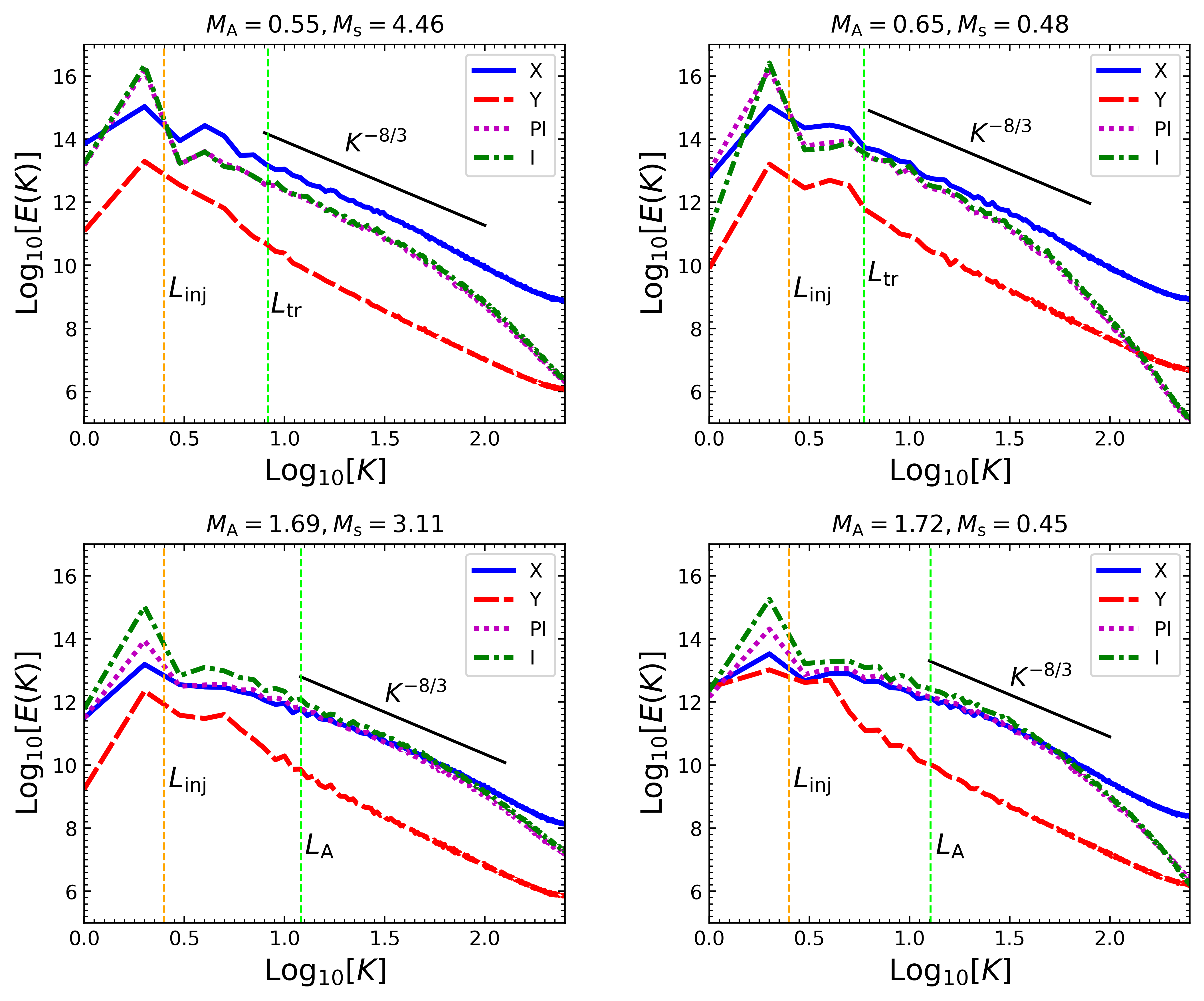}
 \caption{Power spectra of the synchrotron radiation diagnostics: $Q$-$U$ cross ($X$), cross-correlation ($Y$), linear polarization $(PI)$ and total ($I$) intensities in different turbulence regimes. The yellow and green vertical dashed lines represent the injection and transition scales, respectively. }
 \label{fig:power_four}
\end{figure*}

Based on data cubes used in Figure~\ref{fig:two_sf}, we plot the PS of $X$, $Y$, $PI$ and $I$ in Figure~\ref{fig:power_four}. As seen, the scaling indices of PS of $X$ and $Y$ are consistent with $-8/3$ in four turbulence regimes, with an extended inertial range for MHD turbulence. This is a valuable finding in this paper. Due to the presence of numerical dissipation at large wavenumber, the measurements of $PI$ and $I$ can only provide a narrow power-law range, which limits their flexibility to determine the scaling slope of the underlying MHD turbulence. Therefore, the new statistics $X$ and $Y$ have advantages over traditional statistics $PI$ and $I$ in measuring the spectral index and inertial range of MHD turbulence.

In addition, the amplitudes of PS of $X$, $Y$, $PI$, and $I$ in the sub-Alfv{\'e}nic turbulence regime are greater than those in the super-Alfv{\'e}nic turbulence regime. The reason is that large mean magnetic fields for sub-Alfv{\'e}nic turbulence produce more synchrotron radiative information, enhancing Stokes parameters $Q$ and $U$ intensities. In the case of sub-Alfv{\'e}nic turbulence (upper panels), the amplitudes of $X$ are greater than those of $PI$ and $I$ in the inertial range, while it is the opposite in the case of super-Alfv{\'e}nic turbulence (lower panels). Our studies on PS demonstrate that the four statistics explored can measure the scaling slope of MHD turbulence. We emphasize their synergistic measurement abilities to enhance reliability. At the same time, by comparing the amplitudes of different quantities, we can understand their magnetic field strength. It provides a new way to measure magnetization strength, $M_{\rm A}$; further research is necessary in the future.

\subsection{Slope measurement of MHD turbulence with Faraday rotation}

\subsubsection{Effect of radiative frequency}
\label{EFR}
\begin{figure*}
\centering
\includegraphics[width=2\columnwidth,height=0.9\textheight]{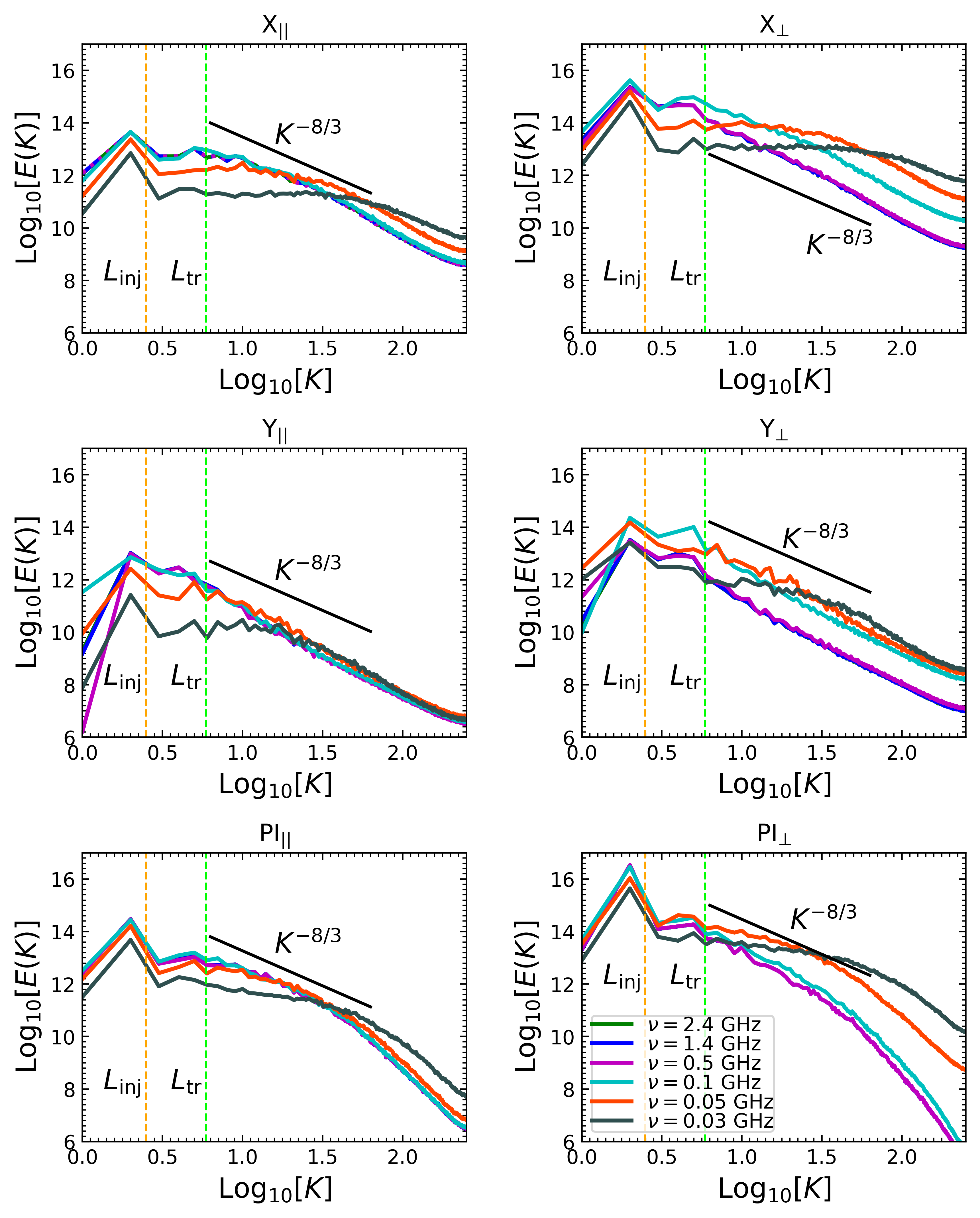}
 \caption{Power spectra of synchrotron radiation diagnostics: $Q$-$U$ cross ($X$), cross-correlation ($Y$) and linear polarization $(PI)$ intensities at different frequencies. Left column: the mean magnetic field is along the LOS. Right column: the mean magnetic field is perpendicular to the LOS. The yellow and green vertical dashed lines are plotted to represent the injection and transition scales, respectively. Our calculations are based on the run3 listed in Table \ref{tab:1}. }
 \label{fig:power_xyp}
\end{figure*}

In this section, we explore how the radiation frequency and the angle between the mean magnetic field and the LOS influence the PS of $X$, $Y$, and $PI$ ($I$ independent of the frequency) in the presence of the Faraday rotation effect. To explore the influence of radiation frequency, we first set the mean magnetic field parallel and perpendicular to the LOS, respectively. Based on the run3 listed in Table~\ref{tab:1}, we show the numerical results in Figure~\ref{fig:power_xyp} for the mean magnetic field parallel (left column) and perpendicular (right column) to the LOS. As is shown in the left column, the PS of $X$, $Y$, and $PI$ follow the scaling law of $-8/3$ in the scales of $<L_{\rm tr}$ for simulations at high frequency (about $\nu \geq 0.1\ \rm GHz$), while in the case of low frequencies, they downward (upward) deviate from $-8/3$ at the small (large) wavenumber regions. This is because, in the high-frequency range, the effect of the Faraday rotation depolarization on the PS is small. With decreasing the frequency, the strong Faraday rotation depolarization leads to a weaker correlation of the radiation signal. As the frequency decreases, the appearance of noise gradually fills the entire synthetical map of Stokes parameters $Q$ and $U$, severely downward distorting the PS statistics at large scales (small wavenumbers). The increase of noise at small scales leads to the upward deviation of the spectral distribution.

The right column of Figure \ref{fig:power_xyp} shows the results of the PS at different frequencies, for which we stress that the mean magnetic field is in the direction perpendicular to the LOS. As seen, the PS of three synchrotron diagnostics $X$, $Y$, and $PI$ satisfy the power law of $-8/3$ at the higher frequencies (about $\nu \geq 0.1\ \rm GHz$) compared with the left panels. At the low-frequency regimes, they show a distribution similar to those of the left column. In addition, we see that the amplitudes of PS of $X$, $Y$, and $PI$ have significant changes at the lower frequencies. In addition to the case of $M_{\rm A}<1$ studied above, we also consider other possible scenarios with $M_{\rm A}>1$ (the relevant results not shown in this paper). When the frequency is higher than 100$~\rm MHz$, the PS of $X$, $Y$, and $PI$ can reveal the spectral index of the underlying MHD turbulence.

As a result, in the case of moderate depolarization, $X$ and $Y$ have more advantages than $PI$ for measuring the scaling slope of MHD turbulence. When the mean magnetic field is parallel to the LOS, it is more helpful to use these statistics to reveal the magnetic turbulence information in the case of a lower frequency (down to about $\nu \simeq 0.1\ \rm GHz$ for our parameter selections).

\subsubsection{Effect of noise and view angle}

\begin{figure*}
\centering
\includegraphics[width=2\columnwidth,height=0.3\textheight]{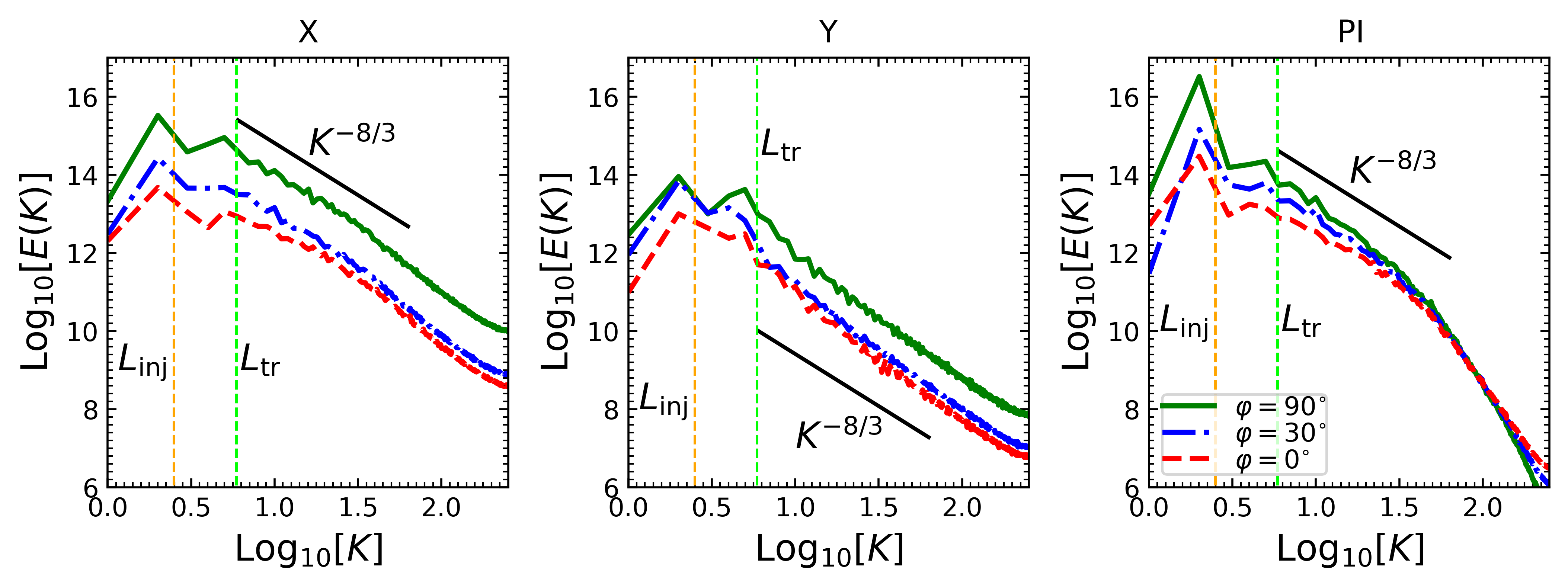}
 \caption{{Power spectra of synchrotron radiation diagnostics: $Q$-$U$ cross ($X$), cross-correlation ($Y$) and linear polarization $(PI)$ intensities at different angles between the mean magnetic field and the LOS at the frequency $\nu$ = 0.14$~\rm GHz$. The yellow and green vertical dashed lines are the injection and transition scales, respectively. Our calculations are based on the run3 listed in Table \ref{tab:1}. }}
 \label{fig:power_theat}
\end{figure*}

 We here explore how the angle between the mean magnetic field and the LOS affects the distribution of PS from relevant diagnostics in Figure \ref{fig:power_theat}, plotted at the frequency of $0.14~\rm GHz$. As is seen in this figure, the inertial range and the amplitude are decreased with decreasing the angle for $X$ and $PI$, resulting in the measured spectral index deviating from -8/3 at a large scale. The reason is that with decreasing angle, the Faraday rotation measure makes synchrotron-polarized signal depolarization. For $Y$, only its amplitude changes rather than its inertial range.

\label{noise_P}
\begin{figure*}
\centering
\includegraphics[width=2\columnwidth,height=0.28\textheight]{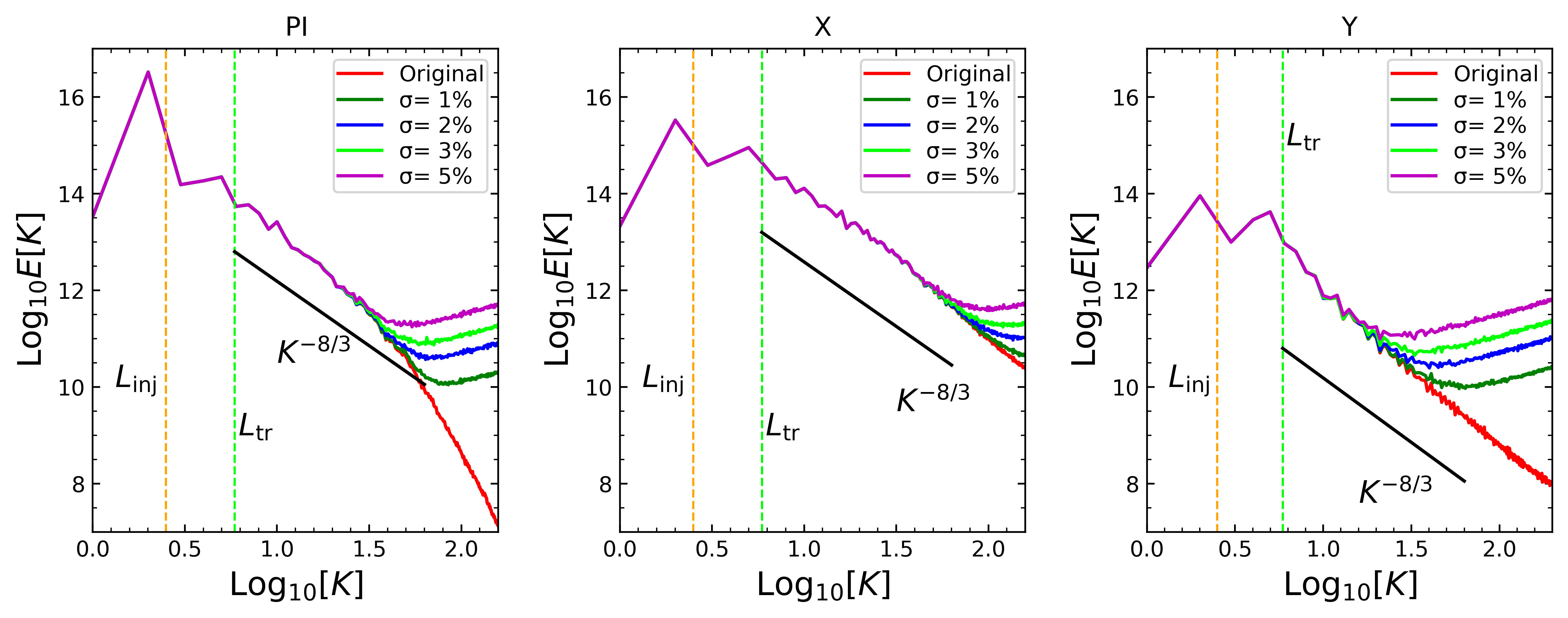}
 \caption{The influence of the noise on the power spectra of synchrotron radiation diagnostics: $Q$-$U$ cross ($X$), cross-correlation ($Y$) and linear polarization $(PI)$ calculated at the frequency $\nu$ = 0.14$\rm~GHz$. The symbol $\sigma$ indicates a standard deviation of Gaussian noise and accounts for a fraction of the mean synchrotron intensities. The yellow and green vertical dashed lines denote the injection and transition scales, respectively.}
\label{fig:noise_ps}
 \end{figure*}

Figure \ref{fig:noise_ps} explores the influence of the noise on the scaling index of PS of $X$, $Y$, and $PI$ at the frequency of $0.14 ~\rm GHz$. We generate a Gaussian noise map with the resolution of $512^2$ and add it to the original image to study the noise effect. Here, the standard deviation of Gaussian noise accounts for the fraction of the mean synchrotron intensity. The figure clearly shows that the PS of $X$, $Y$, and $PI$ adding the Gaussian noise deviate upward from those without noise in the large-$K$ regime. This is because adding noise increases the random fluctuation of the original image, resulting in an increase in the PS in the small-scale region. In addition, we can see that under the same noise level, the inertial range measured by $X$ is wider than those of $PI$ and $Y$. And the higher the level of Gaussian noise, the more obvious the deviation. Therefore, increasing the level of Gaussian noise makes the power-law inertial range narrower.

\subsection{The influence of numerical resolution on the results}

To test the influence of numerical resolution on the measurement of the scaling slope of turbulence, we simulate the data cube with the low numerical resolution of $256^3$ in the same way as the run1 listed in Table \ref{tab:1}. The difference of resolution results in slightly different Mach numbers: $M_{\rm A}=0.70$ and $M_{\rm s}=0.74$ for $256^3$, as well as $M_{\rm A}=0.70$ and $M_{\rm s}=0.87$ for $512^3$.

\begin{figure*}
\centering
 \includegraphics[width=2\columnwidth,height=0.6\textheight]{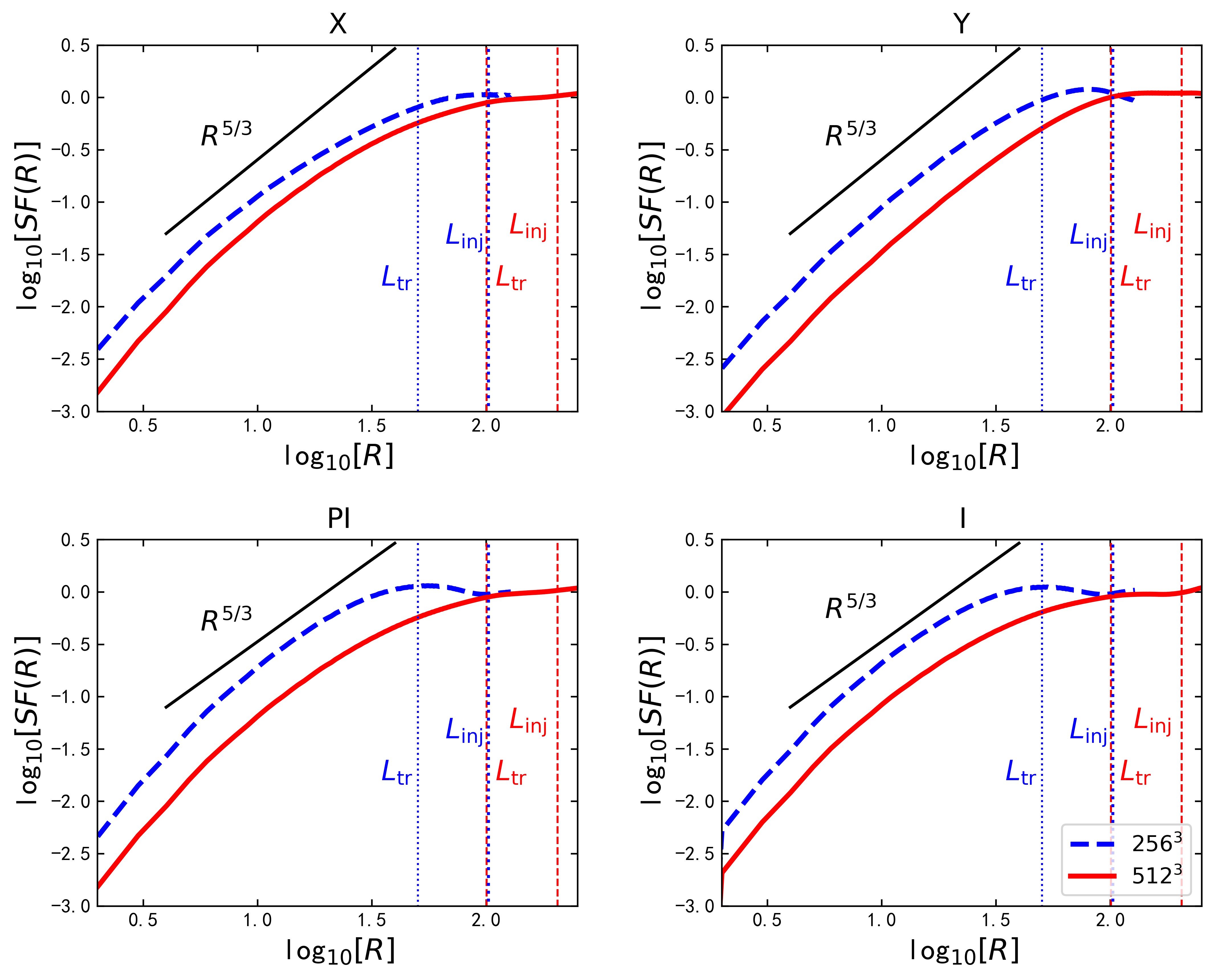}
 \caption{Structure functions of synchrotron radiation diagnostics: $Q$-$U$ cross ($X$), cross-correlation ($Y$), linear polarization $(PI)$ and total ($I$) intensities. The results are presented in two numerical resolutions $256^3$ (blue) and $512^3$ (red), with the run1 listed in Table~\ref{tab:1}. The vertical lines represent the injection and transition scales, respectively.}
\label{fig:resol_sf}
\end{figure*}

Still focusing on the SFs of $X$, $Y$, $PI$ and $I$, we provide numerical results in Figure~\ref{fig:resol_sf}. From this figure, we find that: (1) the measurements from high resolution $512^3$ show better results, namely, closer to $R^{5/3}$, than those from low resolution $256^3$, as expected; (2) For the data cubes with the same numerical resolution, the structure functions of $X$ and $Y$ are more advantageous than those of $PI$ and $I$ in measuring the scaling index and inertial range of MHD turbulence. 

\begin{figure*}
\centering
\includegraphics[width=2\columnwidth,height=0.6\textheight]{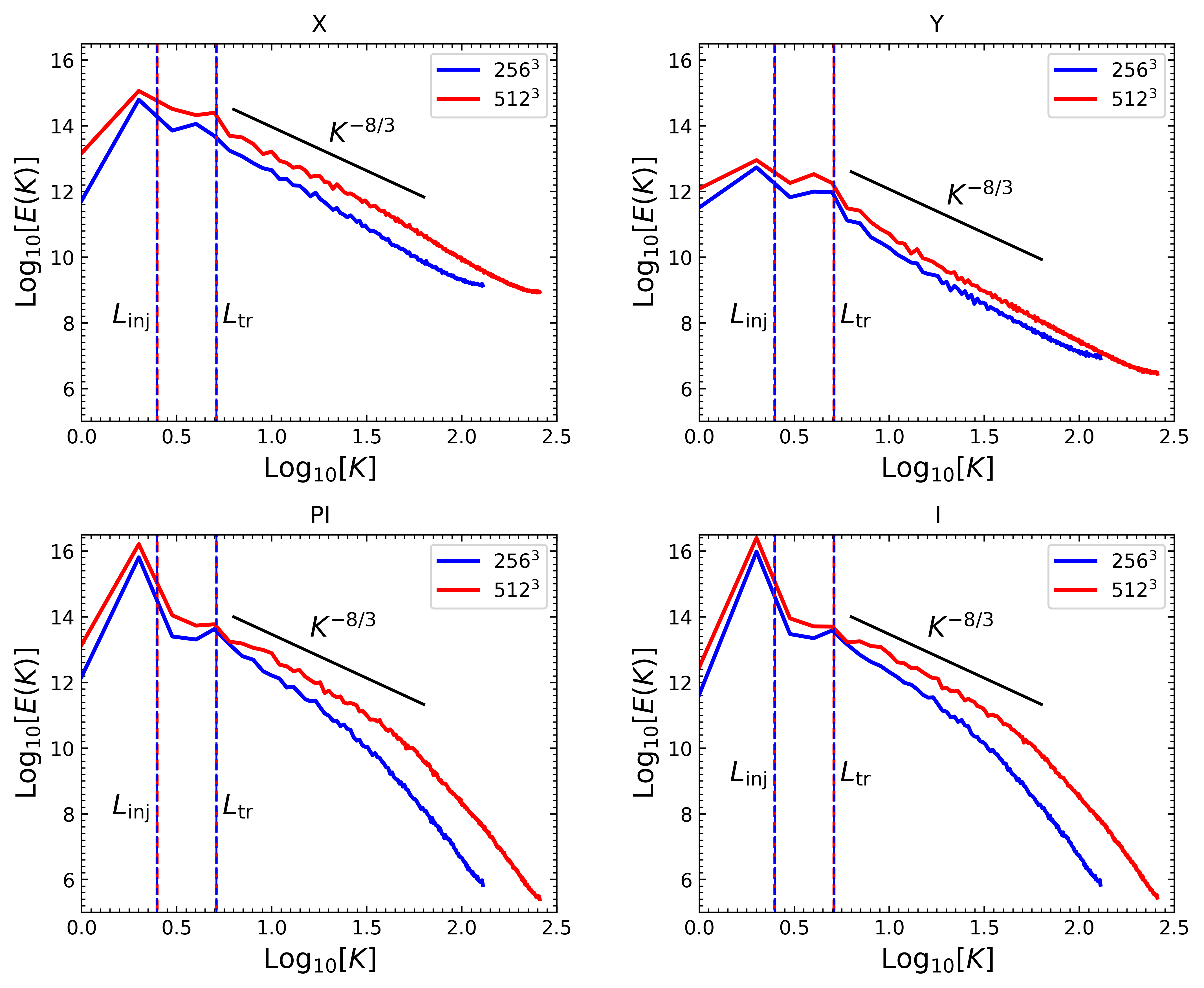}
 \caption{Power spectra of synchrotron radiation diagnostics: $Q$-$U$ cross ($X$), cross-correlation ($Y$), linear polarization $(PI)$ and total ($I$) intensities. The vertical lines denote the injection and transition scales of the underlying MHD turbulence. Numerical results are plotted in resolutions of $256^3$ and $512^3$, using run1 listed in Table~\ref{tab:1}. 
 }
\label{fig:resol_ps}
 \end{figure*}

In addition, we here explore the influence of numerical resolution on the PS of synchrotron radiation diagnostics using two data cubes with the resolutions $256^3$ and $512^3$, the parameters of which correspond to those of data cubes used in Figure \ref{fig:resol_sf}. The numerical results are presented in Figure~\ref{fig:resol_ps}, from which we see that the resolution $512^3$ can better determine the spectral index of $-8/3$ expected in the inertial range. Importantly, we find that the measurements of $Q$-$U$ cross $X$, and cross-correlation $Y$ extend the width of the inertial range, compared with the traditional statistics of linear polarization $PI$ and total $I$ intensities.

Comparing the upper and lower panels of Figure \ref{fig:resol_ps}, we find that the measurements of $PI$ and $I$ have a dissipation at large wavenumbers, i.e., small scales, while $X$ and $Y$ show a weak dissipation to extend the inertial range. Therefore, when recovering the scaling slope of MHD turbulence from observational data, we recommend $X$ and $Y$ statistics. However, we should be particularly cautious that in the largest wavenumber, there is an effect of numerical noise. In this regard, interested readers are advised to refer to \cite{Zhang2016} and \cite{Zhang2018} who found that the noise would cause the spectrum to reverse upwards.

\section{Statistical Results: Decomposition of Compressible MHD turbulence}
\label{results2}
\begin{figure*}
\centering
 \includegraphics[width=2\columnwidth,height=0.3\textheight]{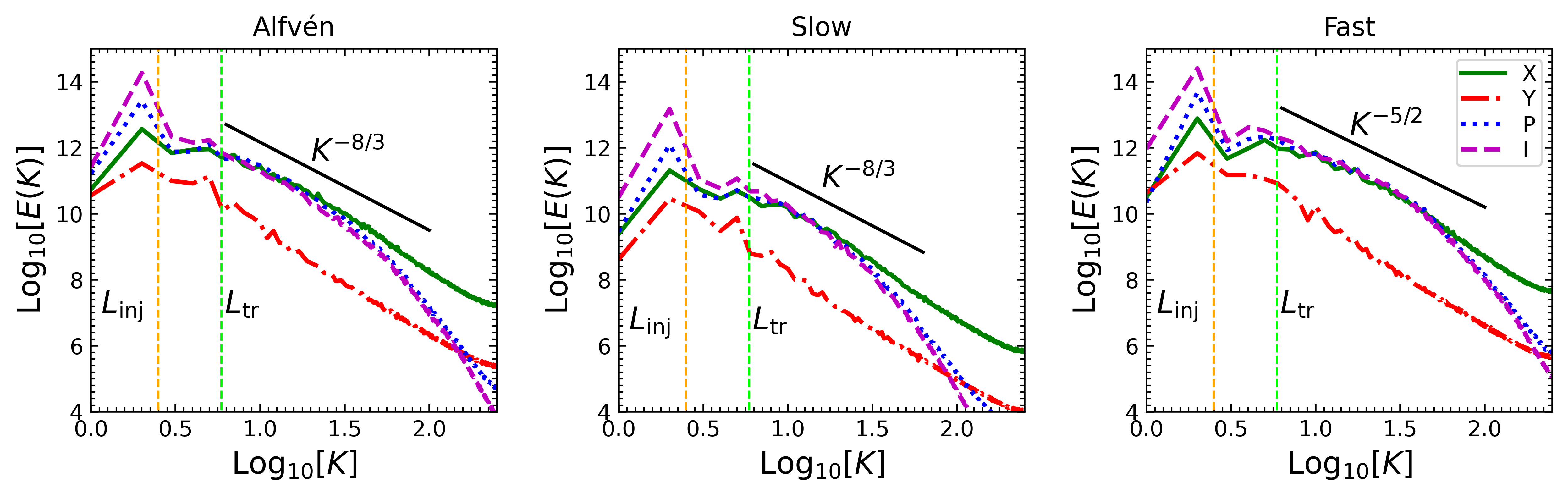}
 \caption{Power spectra of synchrotron radiation diagnostics: $Q$-$U$ cross ($X$), cross-correlation ($Y$), linear polarization $(PI)$ and total ($I$) intensities for Alfv{\'e}n, slow and fast modes. The yellow and green vertical dashed lines represent the injection and transition scales, respectively. The decomposition of data cubes is based on the run3 listed in Table \ref{tab:1}.
 }
 \label{fig:power_mod}
\end{figure*}

With the procedures described in Section \ref{data_produce}, we decompose data cubes of compressible MHD turbulence, namely, run3 listed in Table~\ref{tab:1}, and then explore the PS of $X$, $Y$, $PI$, and $I$ for three modes based on the post decomposition data cubes. As is shown in Figure \ref{fig:power_mod}, the PS of $X$, $Y$, $PI$, and $I$ follows the power law index of $-8/3$ for Alfv{\'e}n and slow modes, and $-5/2$ for fast mode. Meanwhile, it can be seen that the PS of new statistics $X$ and $Y$ has a more extended inertial range than that of traditional $PI$ and $I$ statistics. The properties of PS obtained for these plasma modes are consistent with the theoretical prediction of compressible MHD turbulence described in Section \ref{theo_des}. 

\begin{figure*}
\centering
\includegraphics[width=1\textwidth,height=0.8\textheight]{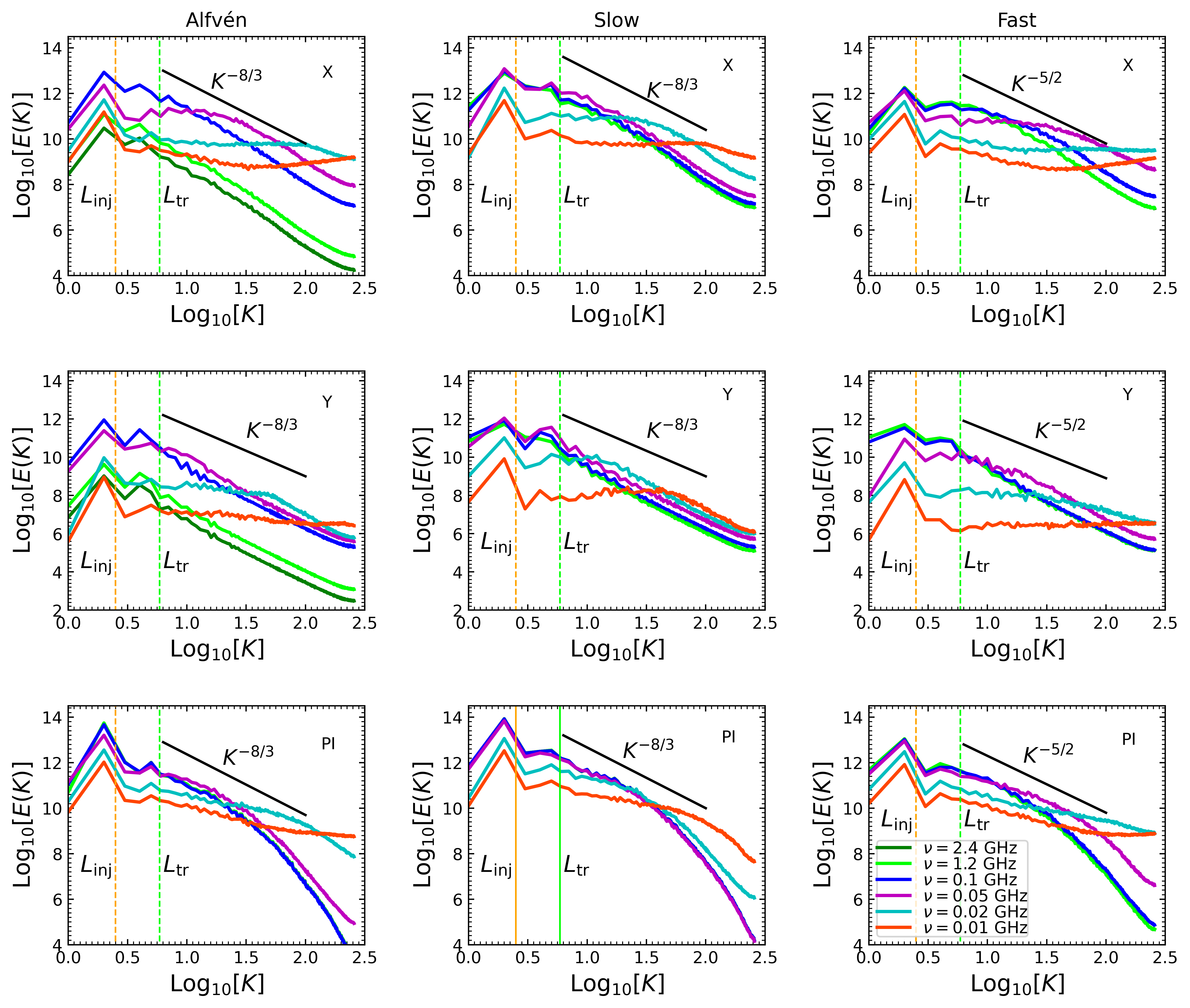}
\caption{Power spectra of synchrotron radiation diagnostics at different frequencies. In the order of the rows: $Q$-$U$ cross $X$ (first row), cross-correlation $Y$ (second), and linear polarization $PI$ (third). In the order of the columns: Alfv{\'e}n (first column), slow (second), and fast (third) modes. The yellow and green vertical dashed lines represent the injection and transition scales, respectively. The decomposition of data cubes is based on the run3 listed in Table \ref{tab:1}. }
\label{fig:power_frmod} 
\end{figure*}

We further explore how the frequency influences the PS of different diagnostics arising from three modes, as shown in Figure~\ref{fig:power_frmod}. Each row corresponds to the PS of the same diagnostic for Alfv{\'e}n (left), slow (middle), and fast (right) modes, respectively, while in each column, the PS of different diagnostics $X$ (upper), $Y$ (middle) and $PI$ (lower) for the same mode. The PS of three diagnostics ($X$, $Y$ and $PI$) follows the scaling slope of $K^{-8/3}$ for Alfv{\'e}n (left column) and slow modes (middle column), and $K^{-5/2}$ for fast mode (right column), at the frequency $\nu> 0.05$ GHz, while they deviate from the expected values ($-8/3$ or $-5/2$) at lower frequencies, particularly, for small wavenumbers. It can be seen that the amplitudes of PS of three diagnostics go up with decreasing frequency in the large-$K$ part. For the slow and fast modes, the physical interpretation of the above phenomena is similar to that of Section \ref{EFR}. But for the Alfv{\'e}n mode, the Stokes parameters of pure Alfv{\'e}n mode at $\varphi
=90^{\circ}$ projects quicker than random walk (\citealt{Lazarian2022}) in the case of low $M_{\rm A}$. Physically it means the Alfv{\'e}n mode without Faraday rotation self-projects to zero if the integration length is large enough. FR destroys this phenomenon and creates fluctuations that are not canceling itself.

Significantly, we find that the PS of various diagnostics arising from three modes depends on the frequency. Note that the PS of different statistics for the slow mode has a slightly weaker dependence on the frequency than the other two modes. The reason is that the slow mode has a small Faraday rotation measure value.

Our research by simulation data confirms that scaling slopes of compressible plasma modes can be obtained from observational data. 
In practice, using observational data to extract the properties of three plasma modes is a challenging subject. Since the scaling index of $-5/2$ for the fast mode is different from the $-8/3$ for the Alfv{\'e}n and slow modes, one could first obtain the properties of the fast mode (\citealt{ZhangHC2020}). However, the more challenging is how to effectively distinguish Alfv{\'e}n and slow modes from the observational data. This deserves more exploratory efforts.

\section{Discussion}
\label{disc}

Synchrotron radiation is an important source of magnetic field information in the interstellar medium environment. Statistics of the total intensity $I$ can provide the information of magnetic field in the plane of the sky. Compared with the intensity $I$, the linear polarization intensity $PI$ can reflect more information about the magnetic field, i.e., not only in the plane of the sky but also in the LOS. The disadvantage is that $P$ can only provide the total polarization information but not the relative importance of $Q$ and $U$, i.e., the relative changes in $Q$ and $U$ values. In this paper, we propose two new diagnostics $X$ and $Y$ together with the traditionally used $PI$ and $I$ to explore the scaling properties of MHD turbulence by PS and SFs and found that PS of $X$ and $Y$ have a larger measurable inertial range compared with $PI$ and $I$. In fact, each technique has its advantages and limitations when measuring turbulence properties. A synergy of various techniques can obtain more comprehensive turbulence information, enhancing the reliability of the turbulence measurement.

Although the PS of synchrotron diagnostics can not provide more information on the spatial structure of MHD turbulence, it is an advantageous statistical method for studying the source, sink and scaling slope of turbulence. As studied in Section~\ref{EFR}, we found that the PS has different amplitudes when the LOS is perpendicular and parallel to the mean magnetic field, so we expect that it can also be an alternative tool for measuring magnetization. 

For compressible MHD turbulence, one can also explore density fluctuation information within MHD turbulence by introducing Faraday rotation. However, the biggest difficulty in measuring magnetic fields through involvement in Faraday rotation studies is that there is not currently a good way to decouple the coupling between vector and scalar.  With a proper understanding of the magnetic field through synergistically related techniques, one can gain insight into the density information. Various MHD turbulence modes have important effects on many astrophysical processes. Therefore, the study of plasma modes is helpful to understand the contribution of different modes to these physical processes, such as the acceleration and diffusion process of cosmic rays (\citealt{Zhang2021,Sampson2023MNRAS}). 

This paper focused on the scaling properties of magnetic fields by PS and SF statistics. Notice that the latter can also be used to recover properties of magnetic field structure and eddy, as done in \cite{Wang2020} and \cite{Zhang2020}. To understand other aspects of MHD turbulence, many other synergistic techniques have been developed based on synchrotron radiation. These techniques include the kurtosis and skewness exploring the anisotropy of MHD turbulence (\citealt{Herron2016}) and constraining the sonic Mach number (\citealt{Burkhart2012}), the quadrupole moment revealing the anisotropy (\citealt{Herron2016, Lee2019, Wang2022}), as well as the gradient statistics measuring magnetic field directions (\citealt{Lazarian2017,Lazarian2018,Zhang2019b, Zhang2019a,Zhang2020,Wang2021,Liu2023MNRAS}) and magnetization (\citealt{Carmo2020,Lazarian2022}). In addition, the PS of the tension force can diagnose the spatial structure of the magnetic structures (\citealt{Schekochihin2004, Waelkens2009, Sun2014}). 

For completeness, our work explored how to get the scaling index of the turbulence by a synchrotron signal in the cases of both subsonic and supersonic turbulence. Indeed, the hot/warm ionized diffuse media with the low $M_{\rm s}$ can be probed by radio synchrotron emission (such as the Galactic ISM with $M_{\rm s}\leq 2$, see \citealt{Gaensler2011}), while some environments still have a large $M_{\rm s}$, such as the regions of active galactic nuclei and supernova remnants interacting with the surrounding cold molecular cloud. Therefore, the $M_{\rm s}$ we explored in this paper were not much greater than 1. For turbulence regimes much larger than 1, one can use alternative approaches, such as velocity channel analysis and velocity correlation spectrum (e.g., \citealt{Lazarian2004a,Yuen2017,Yang2021}). In this work, we did not involve the effect of self-absorption. This process will become important when the magnetic field interacts with relativistic electrons at low-frequency regimes. In the presence of self-absorption, the PS of these statistics may vary not only in the scaling index but also in the inertial range, which provides us with a new research perspective to recover the 3D magnetic field structure.

\section{Summary}
\label{sum}
In this paper, we proposed two new synchrotron diagnostics: the cross intensity $X$ and cross-correlation intensity $Y$ to reveal the MHD turbulence properties. Using their PS and SF together with traditional diagnostics $PI$ and $I$, we have well understood the spectral properties of the underlying compressible MHD turbulence. We focused on exploring how Mach numbers, noise, Faraday depolarization, and numerical resolution affect the spectral measurement of magnetic turbulence. The main results are summarized as follows.

\begin{itemize}
\item The SF of statistics $X$, $Y$, $PI$, and $I$ can determine the scaling slope of MHD turbulence in sub-Alfv{\'e}nic regimes. Interestingly, new statistics $Y$ could better measure the scaling slope compared with other statistics $X$, $PI$, and $I$ in the different Alfv{\'e}nic regimes.

\item The noise does not impede the recovery of the scaling index of MHD turbulence, and the inertial range of PS measured by $X$ is wider than that by $PI$ and $Y$ at the same noise level.

\item In the case of moderate Faraday depolarization, they still improve the scaling slope measurements since the statistics $X$ and $Y$ extend the inertial range. The influence of numerical resolution does not change our conclusions.

\item The change of angle between the mean magnetic field and the LOS does not affect the measurement of the scaling index, but the inertial range and amplitude.

\item Using the synchrotron radiation diagnostics ($X$, $Y$ and $PI$) can measure the spectral properties of Alfv{\'e}n, slow and fast modes. 
\end{itemize}

\section*{ACKNOWLEDGMENTS}
We thank the anonymous referee for valuable comments that significantly improved the quality of the paper. J.F.Z. thanks to the support from the National Natural Science Foundation of China (grant Nos. 11973035), the Hunan Province Innovation Platform and Talent Plan-HuXiang Youth Talent Project (No. 2020RC3045), and the Hunan Natural Science Foundation for Distinguished Young Scholars (No. 2023JJ10039). F.Y.X. acknowledges the support from the Joint Research Funds in Astronomy U2031114 under a cooperative agreement between the National Natural Science Foundation of China and the Chinese Academy of Sciences.

\section*{DATA AVAILABILITY}
The data underlying this paper can be shared on reasonable request to the corresponding author.



\bibliographystyle{mnras}
\bibliography{ms} 




\end{document}